\documentclass[twoside]{article}
\usepackage[accepted]{aistats2023}

\usepackage[round]{natbib}

\usepackage[title]{appendix}
\usepackage{algorithm}
\usepackage{algpseudocode}
\algrenewcommand\algorithmicindent{1.0em}%
\algnewcommand{\LineComment}[1]{\State \(\triangleright\) #1}
\usepackage{amssymb,amsthm}
\usepackage{booktabs}
\usepackage{hyperref}
\hypersetup{ colorlinks, citecolor=blue, linkcolor=blue,  urlcolor=blue}
\usepackage{cleveref}
\usepackage{graphicx}
\usepackage{enumitem} 
\usepackage{makecell}

\crefname{appsection}{Appendix}{Appendices}

\newcommand{\R}{\mathbb{R}}
\newcommand{\PP}{\mathbb{P}}
\newcommand{\E}{\mathbb{E}}
\newcommand{\V}{\mathbb{V}}

\newcommand\Ncal{\mathcal N}
\newcommand\Ccal{\mathcal C}

\newtheorem{proposition}{Proposition}

\newtheorem{assumption}{Assumption}

\newtheorem{?}{Question}

\crefname{assumption}{assumption}{assumptions}

\newcommand{\dd}{\mathrm{d}}

\usepackage{xcolor}

\begin{document}

\runningauthor{L. Riou-Durand, P. Sountsov, J. Vogrinc, C.C. Margossian, S. Power}

\twocolumn[

\aistatstitle{Adaptive Tuning for Metropolis Adjusted Langevin Trajectories}

\aistatsauthor{ Lionel Riou-Durand \And Pavel Sountsov \And  Jure Vogrinc}
\aistatsaddress{ University of Warwick \And  Google Research \And University of Warwick}
\aistatsauthor{ Charles C. Margossian \And  Sam Power }
\aistatsaddress{ Flatiron Institute \And University of Bristol}
]

\begin{abstract}
Hamiltonian Monte Carlo (HMC) is a widely used sampler for continuous probability distributions. In many cases, the underlying Hamiltonian dynamics exhibit a phenomenon of resonance which decreases the efficiency of the algorithm and makes it very sensitive to hyperparameter values. This issue can be tackled efficiently, either via the use of trajectory length randomization (RHMC) or via partial momentum refreshment. The second approach is connected to the kinetic Langevin diffusion, and has been mostly investigated through the use of Generalized HMC (GHMC). However, GHMC induces momentum flips upon rejections causing the sampler to backtrack and waste computational resources. In this work we focus on a recent algorithm bypassing this issue, named Metropolis Adjusted Langevin Trajectories (MALT). We build upon recent strategies for tuning the hyperparameters of RHMC which target a bound on the Effective Sample Size (ESS) and adapt it to MALT, thereby enabling the first user-friendly deployment of this algorithm. We construct a method to optimize a sharper bound on the ESS and reduce the estimator variance. Easily compatible with parallel implementation, the resultant Adaptive MALT algorithm is competitive in terms of ESS rate and hits useful tradeoffs in memory usage when compared to GHMC, RHMC and NUTS.
\end{abstract}
\vfill\null
\section{INTRODUCTION}

Hamiltonian Monte Carlo (HMC; \cite{Duane::1987, neal2011mcmc}) is a widely-known MCMC sampler for distributions with differentiable densities. The use of multiple gradient-informed steps per proposal suppresses random walk behavior and makes HMC particularly powerful for high-dimensional numerical integration. Its implementation in probabilistic-programming languages \citep[e.g.][]{carpenter2017stan,bingham2018pyro,salvatier2016probabilistic,lao2020tfp} leverages automatic differentiation and adaptive tuning with the no-U-turn sampler (NUTS; \cite{hoffman2014nuts}) which require minimal input from the practitioner. 

Robust implementations of HMC rely on randomizing the length of Hamiltonian trajectories \citep[RHMC]{bou2017randomized} to avoid resonances.
NUTS is well established and remains wildly popular but recent studies suggest optimally tuned RHMC samplers can perform better.
Furthermore, NUTS' recursive construction (control-flow-heavy) complicates its implementation when running many parallel chains \citep{radul2020automatically,lao2019unrolled,phan2019iterative}. To take better advantage of modern computing hardware such as GPUs, several schemes for tuning RHMC have been proposed \citep{hoffman2021adaptive,sountsov2021focusing}, showing significant gains of efficiency compared to NUTS.

A second approach to mitigate resonances is based on the kinetic Langevin diffusion. Built upon its discretization, Generalized HMC \citep[GHMC]{horowitz1991generalized} alternates single adjusted steps with partial momentum refreshments. However, the original GHMC algorithm had not been proven competitive, its efficiency being restrained by momentum flips induced upon rejections. To reduce the backtracking caused by these flips, \cite{neal2020non} proposed a slice-sampling approach to cluster the event times of the rejections of a proposal. The efficiency of GHMC was significantly improved by this method, which was later turned into a tuning-free algorithm \citep{hoffman2022tuning}.

In this paper, we focus on a new numerical scheme for approximating the kinetic Langevin diffusion, recently introduced as Metropolis Adjusted Langevin Trajectories \citep[MALT]{riou2022metropolis}. By construction, MALT enables full erasing of the momentum flips and shares several desirable properties with RHMC. Compared to GHMC, some qualitative arguments in favour of MALT were made but no quantitative comparison had been conducted. One objective is to fill this gap.

Our general goal is to develop a tuning-free version of the MALT algorithm, compatible with running many parallel chains. In Section 3 we develop tuning heuristics and algorithmic solutions for each parameter. Built upon these principles, we present in Section 4 an adaptive sampler, applicable for multiple chains as well as for single chain settings. In Section 5 we compare numerically its efficiency to several adaptive samplers (GHMC, RHMC, NUTS) on a set of benchmark models.

To summarize our contributions:
\begin{itemize}
\item We develop tuning heuristics for each of MALT's hyperparameters, building upon recent tuning strategies for RHMC \citep{hoffman2021adaptive,sountsov2021focusing} which target a bound on the Effective Sample Size (ESS). 
\item We propose a method to optimize a sharper bound on the ESS, for which we derive a gradient estimate with reduced variance. This variance reduction method is applicable to both MALT and RHMC and allows for speeding up the adaptation of the trajectory length.
\item We construct an Adaptive MALT algorithm, easily compatible with, but not reliant on, running multiple chains in parallel.
\item We show that Adaptive MALT is competitive in terms of ESS rate and hits useful tradeoffs in memory usage when compared to previous adaptive samplers (GHMC, RHMC, NUTS).
\end{itemize}

\section{BACKGROUND}

This section reviews the use of Hamiltonian dynamics to construct MCMC samplers targeting a differentiable density $\Pi(x) \propto \exp(-\Phi(x))$, for $x\in\R^d$.
We introduce the notion of partial momentum refreshment and present its connection to the kinetic Langevin diffusion,
thereby motivating GHMC.
Unfortunately the utility of GHMC is hindered by momentum flips,
required to preserve detailed balance but statistically inefficient;
we explain how MALT bypasses this issue.
Finally we review existing tuning strategies for HMC and GHMC samplers.

\subsection{Hamiltonian Dynamics}\label{sec:hamiltonian_dynamics}

To describe the motion of a particle $x\in\R^d$ and its velocity $v\in\R^d$ (a.k.a. momentum), Hamiltonian dynamics are defined for a positive definite mass matrix $M\in\R^{d\times d}$ as
\begin{equation}\label{eq:hamiltonian_dynamics}
\dd x_t=M^{-1}v_t\, \dd t ,\qquad\dd v_t=-\nabla\Phi(x_t)\,\dd t.
\end{equation}
By construction, Hamiltonian dynamics preserve the density $\Pi_*(x,v) \propto \exp (-\Phi(x) -  v^\top M^{-1} v/2)$. In general, the ODE (\ref{eq:hamiltonian_dynamics}) does not have a closed form solution. Instead, the leapfrog method is commonly used to integrate this ODE numerically. It is defined for a time-step $h>0$ as the following update $(x_1,v_1)=\textsc{leapfrog}(x_0,v_0;M,h)$ such that
\begin{equation}\label{eq:leapfrog}
\begin{aligned}
    v_{1/2}&=v_0-(h/2)\nabla \Phi(x_0)\\
    x_{1}&=x_0+hM^{-1}v_{1/2}\\
    v_1&=v_{1/2}-(h/2)\nabla \Phi(x_1).
\end{aligned}
\end{equation}
The leapfrog integrator is time-reversible \citep{neal2011mcmc}, i.e. $(x_0,-v_0)=\textsc{leapfrog}(x_1,-v_1;M,h)$. This means that going backwards in time is equivalent to flipping the momenta. HMC is then built upon the following recursion. From a position $x_0\in \R^d$ and a velocity $v_0\sim\Ncal_d(0_d, M)$, a Hamiltonian trajectory of length $\tau>0$ is proposed by composing the leapfrog update for $L=\lceil \tau/h\rceil$ steps. The resultant proposal $(x_L,v_L)$ is faced with an accept-reject test (a.k.a. Metropolis correction) to compensate for the numerical error when solving \eqref{eq:hamiltonian_dynamics}. We note $|v|_{M^{-1}}^2=v^\top M^{-1}v$. With probability $1-\exp(-\Delta^+)$, where $\Delta^+=\max(0,\Delta)$ and
$$\Delta=\Phi(x_L)-\Phi(x_0)+\big(|v_L|_{M^{-1}}^2-|v_0|_{M^{-1}}^2\big)/2,$$
 the proposal is rejected and its momentum is flipped i.e. the output is $(x_0,-v_0)$. Flipping the momentum upon rejection is a technical requirement to ensure that the Metropolis correction preserves $\Pi_*$. Overall, the momentum flips are erased since $v_0$ is fully refreshed at each iteration.

HMC prevents random walk behaviour by proposing trajectories with persistent momentum, thereby exploring the sampling space more efficiently. However, HMC exhibits a phenomenon of resonance when the target $\Pi$ has heterogeneous scales. Typically, fast mixing for one component can result in arbitrarily slow mixing for other components \citep{riou2022metropolis}. This issue makes the efficiency of HMC highly sensitive to the choice of its tuning parameters. 

One solution consists on drawing $\tau$ at random at each iteration (RHMC), thereby smoothing the auto-correlations between successive Hamiltonian trajectories \citep{bou2017randomized}. The choice of the sampling distribution for $\tau$ is further discussed in \Cref{sec:experiments}. In this work, we consider another approach based on partial momentum refreshment.

\subsection{Partial Momentum Refreshment}
\label{sec:partial_momentum_refreshment}

The resonances of Hamiltonian trajectories can be reduced by refreshing partially the velocity along their path. Kinetic Langevin dynamics leverage this property by augmenting (\ref{eq:hamiltonian_dynamics}) with a continuous momentum refreshment induced by a Brownian motion $(B_t)_{t\ge 0}$ on $\R^d$. These are defined for a damping parameter $\gamma\ge 0$ (a.k.a. friction), as
\begin{equation}\label{eq:def:langevin}
\begin{aligned}
    \dd x_t
    &=
    M^{-1}v_t \, \dd t,
    \\
    \dd v_t
    &=
    -\left(\nabla\Phi(x_t)+\gamma v_t\right) \dd t
    +\sqrt{2\gamma} M^{1/2}\dd B_t.
\end{aligned}
\end{equation}

The damping controls the refreshment speed. As $\gamma\rightarrow0$, the SDE (\ref{eq:def:langevin}) reduces to the ODE (\ref{eq:hamiltonian_dynamics}) and exhibits resonances. These resonances vanish as $\gamma$ gets larger. Increasing $\gamma$ too much however decreases the persistence of the momentum and encourages random walk behavior. 
This tradeoff can be solved for a critical choice of damping. For log-concave distributions, quantitative mixing rates have been established by \cite{eberle2019couplings,dalalyan2020sampling, cao2019explicit}. Choosing a mass $M$ in (\ref{eq:def:langevin}) is equivalent to preconditioning the augmented space with the linear map $(x,v)\mapsto (M^{1/2}x,M^{-1/2}v)$. 

Langevin trajectories can be approximated numerically by alternating between the leapfrog update (\ref{eq:leapfrog}) and a partial momentum refreshment, defined for $\eta=e^{-\gamma h}$ as
$$
v’= \eta \, v+\sqrt{1-\eta^2}\, \xi, \qquad \xi\sim\Ncal_d(0_d,M).
$$ 
This partial momentum refreshment leaves $\Pi_*$ invariant. To compensate for the numerical error, one strategy consists in applying the Metropolis correction to each leapfrog step. GHMC is built upon this principle \citep{horowitz1991generalized}, thereby inducing a momentum flip to every rejected leapfrog step; see \Cref{sec:hamiltonian_dynamics}. Unlike HMC, momentum flips are not erased with GHMC, which causes the sampler to backtrack and limits its efficiency \citep{kennedy2001cost}. This also exacerbates the problem of tuning $h$, which must be small enough to prevent momentum flips from becoming overwhelming \citep{bou2010pathwise}.

Several solutions have been proposed to improve GHMC's efficiency. Delayed rejection methods have been used to reduce the number of momentum flips at a higher computational cost \citep{mira2001metropolis,campos2015extra,park2020markov}. We highlight a recent approach based on slice sampling \citep{neal2020non}, which correlates the accept-reject coin tosses over time and encourages larger periods without momentum flips.

\subsection{Metropolis Adjusted Langevin Trajectories}

In this work, we focus on a sampler recently proposed by \cite{riou2022metropolis} to bypass the problem caused by momentum flips. This method consists in applying the Metropolis correction to an entire numerical Langevin trajectory, rather than applying it to each of its $L=\lceil \tau/h\rceil$ leapfrog steps.
The resultant sampler, named Metropolis Adjusted Langevin Trajectories (MALT), is defined by iterating \Cref{alg:malt}. 
\begin{algorithm}
\caption{Metropolis Adjusted Langevin Trajectory}\label{alg:malt}
\begin{algorithmic}[1]
\Function{\textsc{ malt}}{$X;M,\gamma,h,\tau$} 
\State Set $L\gets\lceil \tau/h\rceil$ and $\eta\gets e^{-\gamma h}$ %
\State Draw independent $(\xi_i)_{i=0,\dots,L}\sim\mathcal{N}_d(0_d,M)$ 
\State Set $(x_0,v_0)\gets(X,\xi_0)$ and $\Delta\gets0$
 \For{$i=1$ to $L$}
    \State Refresh $v_{i-1}'\gets\eta\, v_{i-1}+\sqrt{1-\eta^2}\, \xi_i$
	\State Set $(x_i,v_i)\gets\textsc{leapfrog}(x_{i-1},v_{i-1}';M,h)$
\State Update $\Delta\gets \Delta+\big(|v_i|_{M^{-1}}^2-|v_{i-1}'|_{M^{-1}}^2\big)/2$
  
	\EndFor
\State Set $X'\gets x_L$ and $\Delta\gets \Delta+\Phi(x_L)-\Phi(x_{0})$
\State Draw $Z\sim\mathcal{E}xp(1)$
\If{$Z<\Delta$}
    set $X' \gets X$
\EndIf
\State\Return{$X',v_L,x_0,v_0',\Delta$}
\EndFunction
\end{algorithmic}
\end{algorithm}

 In this algorithm, the overall numerical error unfolds as
$
\Delta=\sum_{i=1}^L\delta_i
$,
where 
$$\delta_i=\Phi(x_i)-\Phi(x_{i-1})+\big(|v_i|_{M^{-1}}^2-|v_{i-1}'|_{M^{-1}}^2\big)/2$$ 
is the energy difference incurred by the $i^{th}$ leapfrog step. As a consequence, the probability of accepting an entire trajectory for MALT is higher than the probability of accepting $L$ consecutive leapfrog steps for GHMC. Indeed:
$$
\PP_{\rm GHMC}^{\rm accept}=e^{-\sum_{i=1}^L\delta_i^+} \leq e^{-\left(\sum_{i=1}^L\delta_i\right)^+} =\PP_{\rm MALT}^{\rm accept}.
$$
Detailed balance and ergodicity of \Cref{alg:malt} are proven in \cite{riou2022metropolis}. Notably, the momentum flips get erased by fully refreshing $v_0\sim\Ncal(0_d,M)$ at the start of each trajectory. This ensures reversibility of the sampler with respect to $\Pi$, and makes HMC a special case of MALT obtained for $\gamma=0$. As $\gamma$ gets larger, Langevin trajectories become ergodic, thereby preventing U-turns and resonances. Compared to GHMC, \Cref{alg:malt} also exhibits an advantage for tuning the step-size, which is supported by optimal scaling results derived in \citet{riou2022metropolis}. The tuning of MALT's parameters is further discussed in \Cref{sec:tuning_heuristics}.

\subsection{Adaptive Tuning Strategies}

To measure the efficiency of a MCMC sampler, we consider the problem of estimating $\E[\varphi(x)]$ as $x\sim\Pi$ for a square integrable function $\varphi:\R^d\mapsto\R$. 
The Monte Carlo estimator, $\widehat\varphi_N=N^{-1}\sum_{n=1}^N\varphi(X_n)$, estimates the desired expectation value, with $N$ the number of approximate samples.
A scale-free measurement of $\widehat \varphi_N$'s precision is given by the Effective Sample Size (ESS), defined as
$$
{\rm ESS}_\varphi=\frac{N}{C_\varphi},\qquad C_\varphi= 1+2\sum_{n=1}^\infty{\rm Corr}(\varphi(X_n),\varphi(X_0))
$$
where $X_n$ is the $n^{th}$ iterate of the Markov chain at stationarity. Asymptotically, ${\rm ESS}_\varphi$ measures the number of IID samples required to estimate $\E[\varphi(x)]$ with the same accuracy as $\widehat\varphi_N$; see e.g. \cite{geyer1992practical}. 
To level the computational cost of samplers using multiple gradient steps per iterate, the ESS per gradient evaluation is typically considered as a measure of efficiency.
However, the ESS is often intractable and we cannot directly use it to tune the sampler.  %
The Expected Squared Jumping Distance (ESJD)  generally serves instead as a proxy \citep{pasarica2010adaptively}. It is defined as
$$\mathrm{ESJD}_\varphi = \E \left[ (\varphi(X_1)-\varphi(X_0) )^2 \right].$$
For the ESJD however, choosing a linear re-scaling is not universal, e.g. \cite{wang2013adaptive} optimizes ESJD$/\sqrt{L}$ for RHMC. This question is discussed in \Cref{sec:integration_time}.

For tuning the step-size of HMC, a simple strategy is to target a certain acceptance rate. This strategy is supported by optimal scaling results of \cite{Beskos::2013}. The tuning of the trajectory length is more challenging however. To overcome this problem, \cite{hoffman2014nuts} proposed a stopping rule that stops trajectories before they turn in on themselves, i.e. committing a ``U-Turn". The resultant algorithm (NUTS) requires minimal tuning from the user. NUTS exhibits two limits however: it achieves lower efficiency than optimally tuned RHMC, and the underlying recursion complicates its parallel implementation.  

To alleviate these issues, \cite{hoffman2021adaptive,sountsov2021focusing} have proposed two adaptive schemes (ChEES and SNAPER) aimed at optimizing the (re-scaled) ESJD with a stochastic optimization routine. The choice of $\varphi$ is guided by remarking that RHMC’s efficiency is typically bottlenecked by the squared projection on the principal component. These samplers have demonstrated significant gains of efficiency compared to NUTS, especially in the multi-chain context. 

Partial momentum refreshment received less attention due to the momentum flips issue. Recently though, GHMC’s improvement of \cite{neal2020non} was leveraged to develop an adaptive sampler named MEADS \citep{hoffman2022tuning}. The damping and the step size were tuned by estimating the maximum eigenvalues of $\V(x)$ and $\V(\nabla\Phi(x))$. In the sequel, we build upon recent strategies for tuning RHMC to propose tuning heuristics for MALT.

\section{TUNING HEURISTICS}\label{sec:tuning_heuristics}

In this section, we develop heuristics for tuning each of the following parameters: a diagonal mass matrix $M$, a damping $\gamma$, a time step $h$ and an integration time $\tau$. We support these heuristics with algorithmic solutions, ensuring that computation and memory costs scale linearly with the dimension of the sampling space.

\subsection{Diagonal Preconditioning}
The choice of a preconditioner often plays an important role in the efficiency of gradient-based samplers. A common practice consists of tuning $M$ as proportional to the inverse of the (estimated) covariance matrix of $\Pi$. However this method does not scale linearly with the dimension. We choose a more lightweight solution by tuning a diagonal mass matrix \citep{carpenter2017stan} as
$$M=\max(s)\cdot \text{diag}(s)^{-1}$$
where $s=(s_1,\dots,s_d)$ is composed by estimates of the coordinate-wise variances of $\Pi$. The mass is normalized\footnote{The mass can always be tuned up to a certain constant by rescaling the damping and the time accordingly \citep[Lemma 1]{dalalyan2020sampling}.} by the largest variance to stabilize the learning of the principal component \citep[Appendix C]{sountsov2021focusing}. In practice, diagonal preconditioning often improves the conditioning of the target distribution without reducing it entirely.
Heuristics for the remaining parameters are applied to the pre-conditioned variables $(M^{1/2}x,M^{-1/2}v)$.

\subsection{Damping Parameter}
\label{sec:damping}

The damping is tuned to compensate for the residual conditioning of the target distribution. Choosing the damping requires solving a trade-off between robustness and efficiency. As $\gamma\rightarrow0$, \Cref{alg:malt} yields HMC with fixed $\tau$, which suffers from poor ergodic properties due to the resonance phenomenon; see \Cref{sec:hamiltonian_dynamics}. As $\gamma$ gets large enough, a uniform control of the sampling efficiency becomes possible, typically bottlenecked by the principal component; see \citet{riou2022metropolis}. Taking $\gamma$ too large however induces random walk behavior, mitigating the benefits of the kinetic dynamics. Akin to \cite{hoffman2022tuning}, we set the damping\footnote{Its optimization is discussed for Gaussian targets in \cite{riou2022metropolis}, which advocates a slightly underdamped $\gamma=\lambda^{-1/2}<2\lambda^{-1/2}=\gamma_{\rm critical}$.} as
\begin{align*}
    \gamma=\lambda^{-1/2}
\end{align*}
where $\lambda$ is the largest eigenvalue of $\Sigma$, which we define as the covariance matrix of $M^{1/2}x$. We estimate the principal component and its eigenvalue with an online PCA method known as CCIPCA \citep{weng2003candid}. This method is characterized by the update
\begin{align*}
    w \gets \beta \cdot w + (1 - \beta) \cdot y \cdot (y^\top w)/|w|
\end{align*}
for $y=M^{1/2}(x-m)$ where $x$ is a new sample and $m$ is the mean of $\Pi$. The largest eigenvalue and its eigenvector are respectively estimated by 
$$\lambda=|w|,\qquad z=w/|w|.$$ A formal proof of convergence is established in \cite{zhang2001convergence}. To mitigate the impact of the warm-up phase, we let $\beta = n / (n + a)$ decay with respect to an amnesia parameter $a > 1$. CCIPCA is known to offer a good trade-off between simplicity and performance. Our adaptive scheme is not constrained by this choice though. It is also compatible with methods such as Oja's algorithm or IPCA; see \cite{cardot2018online} for a review.

\subsection{Time Step}\label{sec:time_step}

Guided by optimal scaling results obtained in \citet{riou2022metropolis}, we propose to tune $h$ by targeting an acceptance rate $\alpha^*$ chosen by the user. The simplicity of this strategy gives MALT a significant advantage compared to GHMC, for which the acceptance rate has to be kept close enough to one to avoid overwhelming the Langevin dynamics with momentum flips \citep{bou2010pathwise}. The acceptance probability of \Cref{alg:malt} unfolds as $\alpha(h)=\mathbb{E}[\exp(-\Delta^+)]$, while we aim to find $h$ such that 
$$\alpha(h)-\alpha^*=0.$$
 We leverage the left hand side as a synthetic gradient, for which an unbiased estimator is given by
$$
g_h=\exp(-\Delta^+)-\alpha^*.
$$
The learning of $h$ results from inputting $g_h$ in a stochastic optimization routine called Adam \citep{kingma2014adam}; see \Cref{alg:adam} (gradient ascent).

\begin{algorithm}
\caption{Adaptive moment estimation (Adam)}\label{alg:adam}
\textbf{Hyperparameters: $\delta,\beta_1,\beta_2,\epsilon$ }
\begin{algorithmic}[1]
\Function{\textsc{adam}}{$\theta,m,s,g;n$} 
\State Set $m' \gets \beta_1 m + (1-\beta_1)g$ and $\hat m \gets m'/(1-\beta_1^n)$
\State Set $s' \gets \beta_2 s + (1-\beta_2)g^2$ and $\hat s \gets s'/(1-\beta_2^n)$
\State Update $\theta' \gets \theta+\delta\,  \hat m/(\sqrt{\hat s}+\epsilon)$
\State\Return{$\theta',m',s'$}
\EndFunction
\end{algorithmic}
\end{algorithm}

\subsection{Integration Time}
\label{sec:integration_time}

We propose to tune $\tau$ by focusing on the sampling efficiency of the principal component $z$, measured by $\mathrm{ESS}_\varphi$ for $\varphi(x) = (z^\top M^{1/2} (x-m))^2$. Akin to \cite{sountsov2021focusing}, we consider the variance of the projection rather than its mean to avoid encouraging antithetic behaviors. The computational cost of \Cref{alg:malt} is proportional to the number of gradient evaluations $L=\lceil\tau/h\rceil$, thus increasing $\tau$ is profitable only if the ESS increases at least linearly. A natural criterion to optimize is therefore 
$$\mathcal{C}(\tau)\triangleq\mathrm{ESS}_\varphi (\tau)/\tau.$$
However, deriving an unbiased estimator of the ESS or its gradient is challenging. A common practice is to instead work with the ESJD, which is more suitable for use in stochastic optimization routines such as Adam. 
We remark that the choice of a linear re-scaling for the ESJD is not unanimous in the literature. For instance a re-scaling by $\tau^{1/2}$ was chosen by \cite{wang2013adaptive} for RHMC. In between these two options, we consider a family of criterions indexed by $\rho\in[0,1]$, defined as
\begin{align*}
    \mathcal{C}_\rho(\tau) \triangleq \mathrm{ESJD}_\varphi (\tau) / \tau^{(1 + \rho)/2}.
\end{align*}
The choice $\rho=1$ is supported by the fact that ${\rm ESJD}_\varphi(\tau)$ is proportional to an upper bound on ${\rm ESS}_\varphi(\tau)$ whenever $\rho_\varphi(\tau)\triangleq{\rm Corr}(\varphi(X_{\tau}),\varphi(X_0))\ge 0$ \citep{sountsov2021focusing}. 
MALT being $\Pi$-reversible, we obtain from \cite{geyer1992practical} and \cite{sokal1997monte} that $\Ccal(\tau)$ is upper bounded by the criterion
$$
\overline{\Ccal}(\tau)\triangleq\frac{N}{\tau}\cdot\frac{1-\rho_\varphi(\tau)}{1+\rho_\varphi(\tau)}.
$$
Indeed, if $\rho_\varphi(\tau)\ge 0$ then $\overline{\mathcal{C}}(\tau)\ge N(1-\rho_\varphi(\tau))/\tau$, and ${\rm ESJD_\varphi(\tau)}\propto N(1-\rho_\varphi(\tau))$ implies that $\Ccal_1(\tau)$ is proportional to an upper bound on $\overline{\Ccal}(\tau)$.
In this work, we derive a method to target directly  $$\tau^*=\underset{\tau>0}{{\rm argmax\,}}\overline{\Ccal}(\tau)$$ 
by noticing that $\tau^*$ coincides with the optimizer of $\mathcal{C}_\rho$ when $\rho = \rho_\varphi (\tau^*)$. We therefore propose to target $\mathcal{C}_\rho$ with \Cref{alg:adam} while tuning $\rho$ adaptively as an estimate of the first autocorrelation of $\varphi$ (this heuristic is further supported by a fixed point argument in the supplement).

 Differentiating with respect to $\tau$ yields $\mathcal{C}_\rho'(\tau)=0$ if and only if
\begin{align*}
    \frac{\dd}{\dd \tau}\mathrm{ESJD}_\varphi (\tau)-\frac{1+\rho}{2 \tau}\cdot \mathrm{ESJD}_\varphi (\tau)=0.
\end{align*} 
Considering the solution of the SDE (\ref{eq:def:langevin}) at stationarity, $(\varphi(x_\tau)-\varphi(x_0))^2$ is an unbiased estimator of $\mathrm{ESJD}_\varphi (\tau)$.
In addition, since $\dd x_\tau=M^{-1}v_\tau \dd \tau$, an unbiased estimator of the derivative of $\mathrm{ESJD}_\varphi(\tau)$ is given by
$$
    \delta(x_\tau,x_0,v_\tau) \triangleq 2\, (\nabla \varphi(x_\tau)^\top M^{-1}v_\tau)\cdot (\varphi(x_\tau)-\varphi(x_0)) 
$$
where $\nabla\varphi(x)=2\,  (z^\top M^{1/2}(x-m))\cdot M^{1/2}z$. We leverage the time reversibility of (\ref{eq:def:langevin}) and remark that $(x_\tau,x_0,v_\tau)$ and $(x_0,x_\tau,-v_0)$ have the same distribution. This enables the use of a reduced variance estimator, defined as
$$
g ( x_\tau, v_\tau, x_0, v_0 )\triangleq\frac12\Big( \delta(x_\tau,x_0,v_\tau) + \delta(x_0,x_\tau, - v_0) \Big).
$$
The two components of this averaged estimator are asymptotically uncorrelated as $\tau\rightarrow\infty$ (see supplement). Therefore, considering $g$ in place of $\delta$ is expected to reduce the variance by a factor $2$ as $\tau$ gets large . This variance reduction method is applied to both MALT and RHMC in the experiments of \Cref{sec:experiments}, to speed up the adaptation. As an example of this effect for MALT on the Hierarchical Linear Model, the variance of the $\tau$ estimate goes from 0.11 to 0.057 when the improved estimator is used.

\section{ADAPTATION WITH MANY CHAINS}

\begin{figure}[!ht]
  \centering
    \includegraphics[scale=0.70]{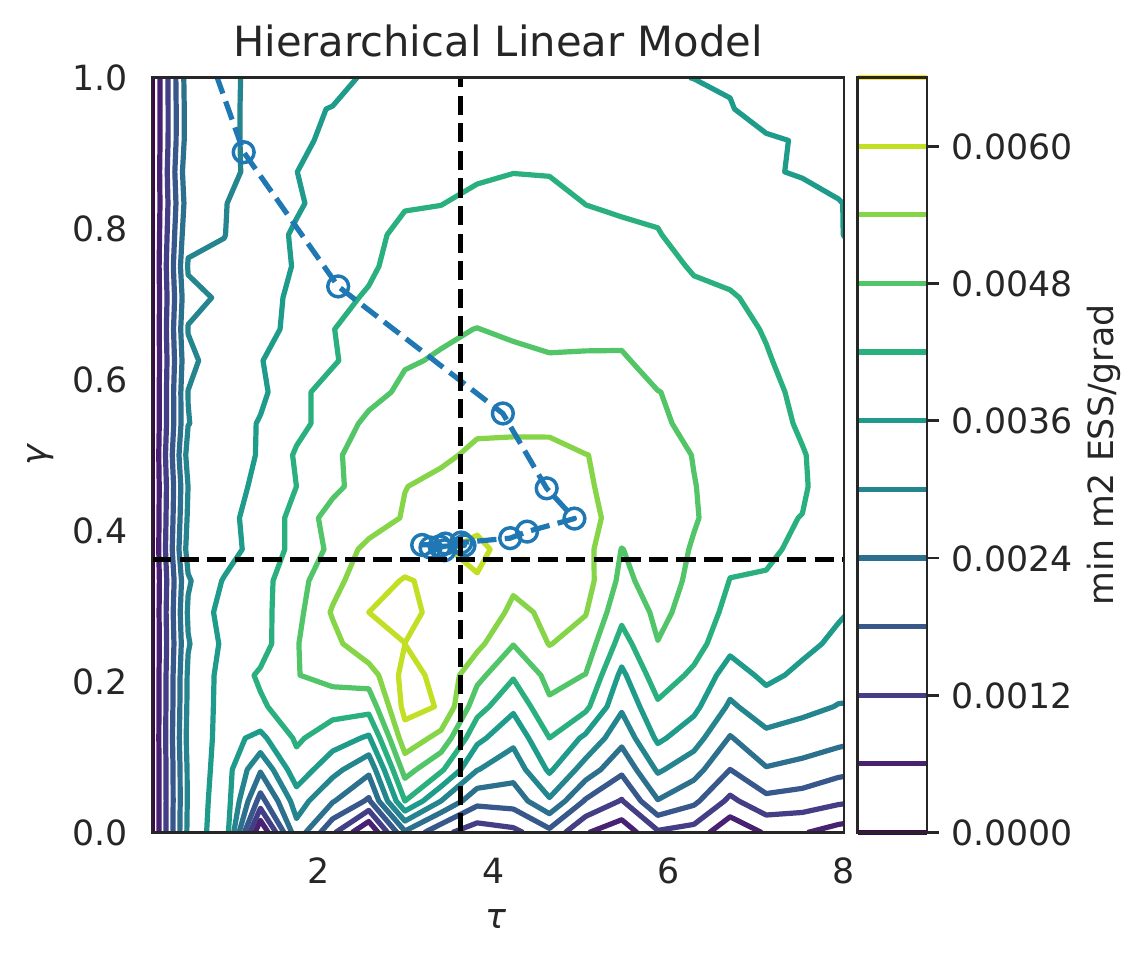}
\caption{
    Trace of the $\tau$ and $\gamma$ hyperparameters from one run of Adaptive MALT on the Hierarchical Linear Model; see \Cref{sec:experiments}. Every dot represents 5 steps of the algorithm. In the background, the surface plot is the minimum (across dimensions) ESS/grad for estimating the centered second moment, computed via grid search.
}
    \label{fig:evolution}
\end{figure}

We now describe an Adaptive MCMC algorithm for tuning the hyperparameters of MALT which combines the heuristics described in the previous section. We assume that we can run $K$ chains in parallel, which allows us to share statistical strength between them as well as being amenable to modern SIMD hardware accelerators. The heuristics discussed previously are readily adapted to this setting. 
\Cref{alg:adaptive_malt} outlines the resultant \emph{Adaptive MALT}.
The moments $m,s,w,m_2,s_2$ required by the optimization of $\tau$ and the tuning of $M$ and $\gamma$ are updated with respect to the functions $f_\psi(x;\theta)$ summarized in \Cref{tab:param_updates}. 
\Cref{fig:evolution} shows an example of the adaptation trace produced by Adaptive MALT.

\begin{algorithm}
\caption{Adaptive MALT} \label{alg:adaptive_malt}
\textbf{Updates:} Parallel states: $X_1,\dots,X_K$. Online estimates of  $\theta=(m,s,w,m_2,s_2)$ and
$h,m_h,s_h,\tau,m_\tau,s_\tau,c$.\\
\textbf{Hyperparameters:} $N_{\text{adapt}},N_{\text{clip}},a$
\begin{algorithmic}[1]
\For{$n=1$ to $N_{\text{adapt}}$}
\LineComment{Compute the mass matrix and friction.}
\State set $M\gets\max(s)\cdot \text{diag}(s)^{-1}$ and $\gamma\gets|w|^{-1/2}$
\If{$n\le N_{\text{clip}}$} set $\tau\gets h$ 
\EndIf
\State set $\beta\gets n/(n+a)$ and $\rho\gets c^+/s_2$
\LineComment{Perform MALT updates.}
 \For{$k=1$ to $K$}\textbf{ in parallel}
    \State $X_k,v_\tau,x_0,v_0,\Delta_k\gets\textsc{malt}(X_k;M,\gamma,h,\tau)$
        \State $g_k\gets g(X_k,v_\tau,x_0,v_0)- \frac{1+\rho}{2\tau}(\varphi(X_k)-\varphi(x_0))^2$
        \State $c_k\gets (\varphi(X_k)-m_2)(\varphi(x_0)-m_2)$
	\EndFor
        \LineComment{Adapt step size.}
		\State $g_h\gets \frac{1}{K}\sum_{k}e^{-\Delta_k^+}-\alpha^*$
	\State $\log(h),m_h,s_h\gets\textsc{adam}(\log(h),m_h,s_h,g_h;n)$
        \LineComment{Adapt trajectory length.}
		\State $g_\tau\gets  \frac{1}{K}\sum_{k} g_k$
	\State $\log(\tau),m_\tau,s_\tau\gets\textsc{adam}(\log(\tau),m_\tau,s_\tau,g_\tau;n)$
        \LineComment{Adapt online estimates.}
	\State $c\gets \beta c+ (1-\beta)\frac{1}{K}\sum_{k}c_k$
 \For{$\psi\in\{m,s,w,m_2,s_2\}$}
    \State $\psi\gets \beta \psi+(1-\beta)\frac{1}{K}\sum_{k}f_\psi(X_k;\theta)$
	\EndFor
	\EndFor
\end{algorithmic}
\end{algorithm}

\begin{table}[!h]
\caption{Summary of the updates for $y_x=M^{1/2}(x-m)$, $M=\max(s)\cdot \text{diag}(s)^{-1}$ and $z=w/|w|$ in \Cref{alg:adaptive_malt}.}
    \label{tab:param_updates}
\begin{center}
    \begin{tabular}{c|c|c}
      \text{Parameter}&$\psi$  & $f_\psi(x;\theta)$ \\
      \hline
       \text{Vector of means}&  $m$ & $x$\\
     \text{Vector of variances}& $s$ & $(x-m)^{\odot 2}$\\
       \text{ Principal eigenvector}&   $w$ & $\big(z^\top y_x \big)\cdot y_x$\\
     \text{ Mean squared projection} & $m_2$ & $\big(z^\top y_x\big)^2$\\
      \text{ Variance squared projection} &$s_2$ & $\big(\big(z^\top y_x\big)^2-m_2\big)^2$  \\
    \end{tabular}
\end{center}
\end{table}

We consider here a finite adaptation scheme, for which ergodicity is ensured by the ergodicity of each MALT kernel; see \citet[Proposition 2]{roberts2007coupling}.
In \Cref{alg:adaptive_malt}, the time step and the integration time are optimized by inputting the averaged gradient estimates $g_h$ and $g_\tau$ into Adam, defined in \Cref{alg:adam}. 
The $\tau$ optimization targets the criterion $\Ccal_\rho$ by setting $\rho$ adaptively as described in \Cref{sec:integration_time}. This adaptation can be turned off by setting $\rho=1$, in which case the estimation of $m_2$ and $s_2$ is no longer required; see \Cref{sec:experiments} for a comparison. 
The updates we present are averaged across chains using an arithmetic mean to preserve unbiasedness for $g_h$ and $g_\tau$. For robustness, a common practice is to use instead the harmonic mean \citep{hoffman2021adaptive} to average the acceptance rates across chains when tuning the step size.

While we frame the Adaptive MALT algorithm for the case where $K > 1$, it is defined to remain functional even in cases where $K = 1$. This is a setting where there is insufficient memory to run multiple chains in parallel, or a hardware accelerator is not available, or already well-utilized by SIMD computation within $\Phi$ and $\nabla \Phi$. We observe that setting $K = 1$ requires lower learning rates and longer chains to allow for the hyperparameters to converge.

It is interesting to compare the implementation of many-chain MALT and RHMC. By construction, MALT uses a single integration time $\tau$ while RHMC needs to randomize it every iteration. A straightforward implementation of many chain RHMC would require each iteration to simulate the number of leapfrog steps necessary for the chain that has the longest integration time draw, causing wasted computation for the rest of the chains. Practical implementation of RHMC \citep{hoffman2021adaptive,sountsov2021focusing} instead choose to share the random $\tau$ draw across chains, trading efficiency for correlation across chains. In contrast, many-chain MALT does not need to make similar trade-offs.

When comparing MALT and GHMC as implemented by MEADS, one key difference is the estimator of largest eigenvalue $\lambda$. MALT, as discussed in \Cref{sec:damping} uses CCIPCA which has compute complexity of $O(Kd)$. MEADS, due to its non-adaptive nature, requires the use of an estimator that can return good estimates of $\lambda$ using only the current state of the MCMC chain. MEADS uses an estimator based on approximating the ratio $\mathrm{tr}({\Sigma^2}) / \mathrm{tr}(\Sigma) \approx \lambda$ with an algorithm with a compute complexity of $O(K^2 d)$. For large number of chains, the additional computational burden can be significant, which is compounded by the fact that MEADS requires a large number of chains to provide accurate estimates whereas Adaptive MALT does not.

\section{EXPERIMENTS}\label{sec:experiments}

\begin{table*}[!htb]

\caption{
Bayesian models used for the experiments. The code snippets refer to the Python class invocation to construct the model (the classes are all found in the {\small\texttt{inference\_gym.targets}} module).
\vspace{0.2cm}
}
\label{tab:models}

\centering
\begin{tabular}{ l|c|l }
 Model & $d$ & Inference Gym Snippet \\
\hline
Logistic Regression & 25 & \scriptsize\texttt{GermanCreditNumericLogisticRegression()} \\
Brownian Bridge & 32 & \scriptsize\texttt{BrownianMotionUnknownScalesMissingMiddleObservations(use\_markov\_chain=True)} \\
Sparse Logistic Regression & 51 & \scriptsize\texttt{GermanCreditNumericSparseLogisticRegression(positive\_constraint\_fn="softplus")} \\
Hierarchical Linear Model & 97 & \scriptsize\texttt{RadonContextualEffectsIndiana()} \\
Item Response Theory & 501 & \scriptsize\texttt{SyntheticItemResponseTheory()} \\
Stochastic Volatility & 2519 & \scriptsize\texttt{VectorizedStochasticVolatilitySP500()} \\
\end{tabular}

\end{table*}

\begin{figure*}[!tb]
  \centering
    \includegraphics[scale=0.65]{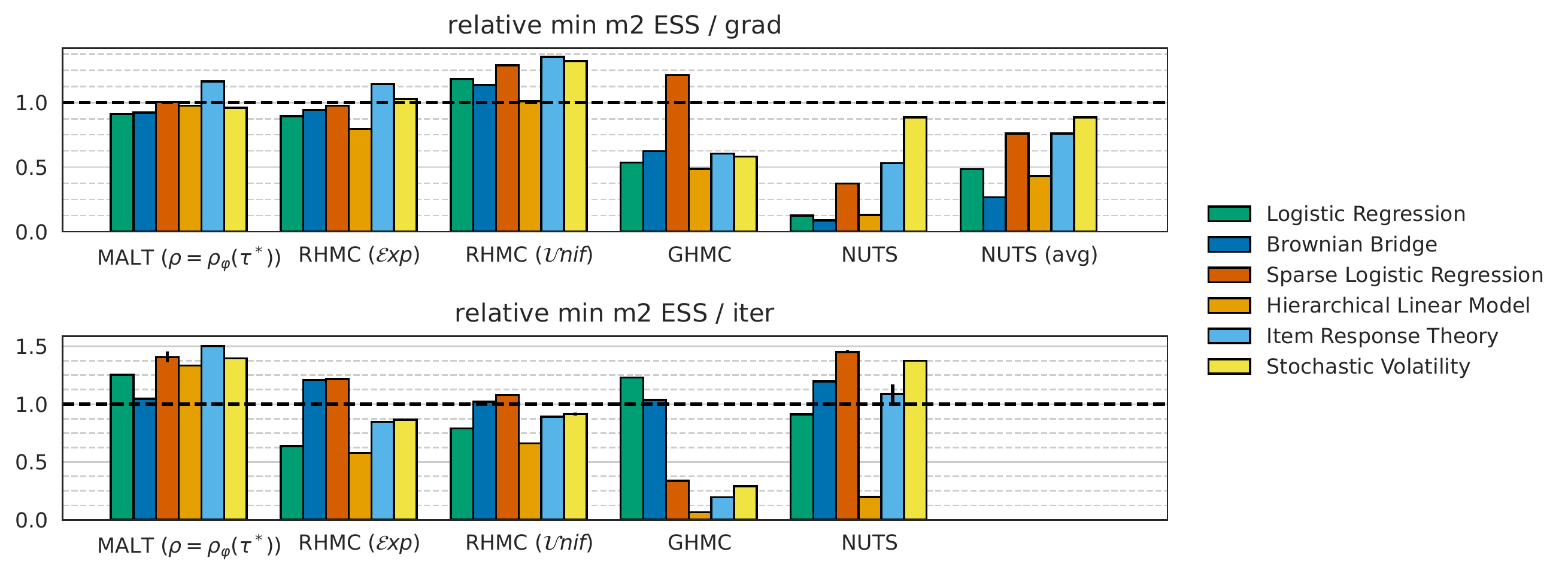}
\caption{
    Minimum ESS/grad and ESS/iter for estimating the centered second moment across algorithms and models. All values are normalized by the corresponding efficiency measure attained by Adaptive MALT with fixed $\rho = 1$. Values are 10$^\text{th}$ percentile across 20 runs. Errors bars are 3 SEM estimated via bootstrap, but are often too small to see.
}
    \label{fig:ess}
\end{figure*}

We validate Adaptive MALT and compare it to a number of baselines by targeting the posterior distributions of several Bayesian models. All the models are sourced from the Inference Gym \citep{sountsov2020inference}, and are summarized in \Cref{tab:models}. In addition to Adaptive MALT, we also examine the performance of RHMC (as implemented by SNAPER; \cite{sountsov2021focusing}), GHMC (as implemented by MEADS; \cite{hoffman2022tuning}) and NUTS (following the implementation from \cite{sountsov2021focusing}). For MALT we examine the effect of tuning $\rho$ or keeping it fixed at 1 (see \Cref{sec:integration_time}). We use the performance of the latter configuration to normalize the results from the remaining algorithms. For RHMC we sample the trajectory lengths from a uniform distribution (\emph{RHMC (}$\mathcal{U}nif$\emph{)}) or an exponential distribution (\emph{RHMC (}$\mathcal{E}xp$\emph{)}), both with mean $\tau$. Additionally, we use the improved estimator for computing the trajectory criterion gradient as described in \Cref{sec:integration_time}. The target acceptance rate is chosen as $\alpha^*=80\%$, slightly above the asymptotically optimal 65\% rate; see \cite{Beskos::2013} for HMC and \cite{riou2022metropolis} for MALT. This is a common practice to ensure robustness with minimal loss of efficiency for HMC \citep{betancourt2014optimizing}. A detailed study of this tradeoff for MALT is beyond our scope.

We run all algorithms for 128 parallel chains, with 5000 adaptive warmup iterations, followed by 400 non-adaptive warmup iterations, and finally 1600 sampling iterations used for the collection of statistics. We repeat each simulation run 20 times with different PRNG seeds.
All algorithms are implemented using the Python package JAX \citep{jax2018github} and run on a GPU\footnote{Implementation of the experiments is available at \url{https://github.com/tensorflow/probability/tree/main/discussion/adaptive_malt}}.

The number of chains is chosen large enough to meet MEADS’ requirements for its parameter tuning scheme. The MEADS implementation is also sensitive to initialization (in the original work, a MAP estimate is used), so to level the playing field we initialize all algorithms with samples derived from Automatic Differentiation Variational Inference (ADVI) \citep{kucukelbir2016automatic}.

We report the minimum Effective Sample Size (ESS) of the sampling phase among the $d$ squared components. The latter is relevant for estimating the marginal variances of the posterior distribution. Notably, this metric is independent from the estimation of the principal component, and accounts for resonances (see \Cref{sec:damping}). The ESS is normalized either by the number of $\nabla \Phi$ evaluations (ESS/grad), or the number of MCMC iterations (ESS/iter). The first metric reflects the efficiency per unit of computation, as gradient evaluations typically dominate the computation time in high dimension. The second metric reflects the efficiency per unit of storage, focusing on situations when memory is at a premium, and many iterations must be stored for analysis.
Such a situation can arise when fitting a high-dimensional model using a large number of chains.

There is a subtlety when computing ESS/grad for NUTS in a parallel-chain setting, as its ragged computation will necessarily waste some computation compared to the other algorithms. To address this we compute the number of gradients as both the maximum across chains for each step (reflecting the actual implementation) and the mean across chains (labeled as \emph{NUTS (avg)}). The latter measures performance that is closer to the non-SIMD single chain implementation, but is difficult to reach in a SIMD setting. For MEADS, we follow \cite{hoffman2022tuning} and thin the chain by a factor of 10, meaning that each iteration consumes 10 gradient evaluations. Both metrics ESS/grad and ESS/iter are sensitive to the choice of thinning, but tuning it optimally is outside the scope of this work.

The results are summarized in \Cref{fig:ess}.

When considering ESS/grad, Adaptive MALT is competitive with the best algorithms. It outperforms MEADS (GHMC) in 5 out of the 6 problems. It outperforms NUTS in all 6 problems. However, it is often slightly less efficient than RHMC with uniform jitter. Aside from examining MALT's performance, our experiments also highlight one hitherto under-explored design choice of RHMC: the distribution of trajectory lengths. We observe that the uniform distribution produces a higher efficiency sampler than the exponential distribution, which has been previously studied for theoretical convenience \citep{riou2022metropolis, bou2017randomized}. A possible explanation for this is that integration times sampled from an exponential distribution which exceed $2 \tau$ are wasteful due to performing a U-turn. Compared to setting $\rho=1$, we observe that targeting $\rho=\rho_\varphi(\tau^*)$ only improves the efficiency in one problem: the Item Response Theory model.  Despite the fact that adapting $\rho$ targets a sharper upper bound on the ESS (in continuous time), it also induces a smaller step-size as the trajectory length increases. As a result, the numerical efficiency is optimized by choosing slightly shorter trajectories.

When looking at ESS/iter, Adaptive MALT performs better than the other algorithms in most problems. This is mainly due to the fact that MALT prefers slightly longer trajectories than RHMC. This is emphasized when setting $\rho$ adaptively. Comparing both algorithms based on Langevin dynamics, we see that MALT often outperforms MEADS (GHMC), all the while not requiring choosing a thinning schedule and being amenable to few-chain adaptation.

\section{DISCUSSION}

This paper proposes tuning guidelines for each of MALT's hyperparameters. The tuning of MALT leverages recent strategies for tuning RHMC \citep{hoffman2021adaptive,sountsov2021focusing}, consolidated with: (i) the use of a parameter-free online PCA method (CCIPCA), to estimate the principal component and its eigenvalue, (ii) an adaptive re-scaling of the ESJD which targets a sharper upper bound on the ESS, (iii) a reduced variance estimator of the gradient of $\tau$. These methods are new and applicable beyond the scope of MALT, e.g. (iii) enables a faster adaptation of RHMC in \Cref{sec:experiments}. 

The resulting Adaptive MALT is user-friendly, readily applicable to a broad range of targets and benefits from a high-performance implementation in JAX.
Unlike NUTS, it is amiable to hardware accelerators and can be run efficiently using many chains in parallel.
The benefits of running many chains include better adaptation (as discussed in this paper and references)
but also opportunities to make sampling phases arbitrarily short and construct more principled convergence diagnostics \citep{margossian2022nrhat}.
With the many chains regime becoming more standard, MALT's effective use of memory is particularly compelling.

Compared to GHMC, the MALT algorithm bypasses the momentum-flips problem by applying the Metropolis correction to entire Langevin trajectories. Our work validates this qualitative advantage by showing that Adaptive MALT outperforms GHMC numerically, even when improved by a slice-sampler \citep{neal2020non, hoffman2022tuning}. MALT’s ability to avoid momentum flips paves the way for developing new efficient algorithms, e.g. based on other types of partial momentum refreshment.

\acknowledgments

We thank Matthew D. Hoffman as well as the anonymous reviewers for their helpful comments and
suggestions. LRD was supported by the EPSRC grant EP/R034710/1. JV was supported by the EPSRC grant EP/T004134/1.
SP was supported by the EPSRC grant EP/R018561/1.

\bibliographystyle{apalike}
\bibliography{tuning_free}

\clearpage

\begin{appendices}

\onecolumn

{\centering
  {\Large\bfseries Supplementary Material of ``Adaptive Tuning for Metropolis Adjusted Langevin Trajectories" \par}}
 \bottomtitlebar

\crefalias{section}{appsection}

\section{ADAPTIVE REGULARIZATION OF THE ESJD}
In this section, we derive the adaptive regularization of the ESJD presented in \Cref{sec:integration_time}. 
We note $(x_\tau,v_\tau)_{\tau\ge 0}$ the stationary solution of the Langevin SDE defined in \Cref{sec:partial_momentum_refreshment}. We consider the criterions $\overline{\mathcal{C}}(\tau)$ and $\mathcal{C}_\rho(\tau)$ for $\rho\in[0,1]$, defined in \Cref{sec:integration_time}. Our analysis holds for any continuously differentiable function $\varphi:\R^d\mapsto\R$ such that each expectation of this section exists.
We make repeated use of the short-hand notations
$$
J(\tau)\triangleq\E\big[(\varphi(X_\tau)-\varphi(X_0))^2\big], \qquad \rho(\tau)\triangleq{\rm Corr}\big(\varphi(X_\tau),\varphi(X_0)\big).
$$
We highlight that $\rho\in[0,1]$ denotes a tuning parameter, while $\rho(\tau)$ stands for the auto-correlation function after time $\tau$. Since $J(\tau)\propto1-\rho(\tau)$, we get
\[
\overline{\mathcal{C}}(\tau)
\propto
\frac{J(\tau)}{\tau\,(1+\rho(\tau))},\qquad \mathcal{C}_\rho(\tau)
=
\frac{J(\tau)}{\tau^{(1+\rho)/2}}.
\]

\begin{assumption}\label{ass:unique_maximizers}
The criterions $\overline\Ccal$ and $\Ccal_\rho$ have unique maximizers for every $\rho\in[0,1]$, denoted
$$
\tau^*\triangleq\underset{\tau>0}{{\rm argmax}}\, \overline\Ccal (\tau),\qquad \tau_\rho^*\triangleq\underset{\tau>0}{{\rm argmax}}\, \Ccal_\rho(\tau).
$$
\end{assumption}

\begin{assumption}\label{ass:decreasing_rho}
The map $\tau\mapsto \rho(\tau)$ is positive decreasing on $[\tau_1^*,\tau_0^*]$.
\end{assumption}

On Gaussian targets, these two conditions are satisfied for the map $\varphi(x)=(z^\top M^{1/2}(x-m))^2$ considered in \Cref{sec:integration_time}. In such a case, the criterions $\overline\Ccal(\tau)$ and $\Ccal(\tau)={\rm ESS}_\varphi(\tau)/\tau$ also coincide, due to the geometric decay of the autocorrelations.\newline

For any $\tau>0$, the auto-correlation $\rho(\tau)$ can be approximated with a long run of MALT. Conversely for any $\rho\in[0,1]$, the optimizer of $\Ccal_\rho$ can be approximated from a long run of Adam. In \Cref{sec:integration_time}, we propose to target adaptively $\Ccal_\rho$ with Adam while setting $\rho$ equal to the auto-correlation of the MALT chain.
We support this strategy with a fixed point argument in the next proposition.\newline

\begin{proposition}\label{prop:fixed_point}
Suppose \Cref{ass:unique_maximizers,ass:decreasing_rho} hold. Define $f(s)\triangleq \tau^*_{\rho(s)}$  for any $s\ge0$ such that $\rho(s)\ge 0$. Then $\tau^*$ is the unique attracting fixed point of $f$. Furthermore, the sequence $s_{k+1}=f(s_k)$ converges to $\tau^*$ for any start $s_0\in [\tau_1^*,\tau_0^*]$.
\end{proposition}

The claim of \Cref{prop:fixed_point} is established by proving the following properties:
\begin{enumerate}[label=(\roman*)]
    \item $\tau^*$ is the unique fixed point of $f$.
    \item $\rho\mapsto \tau_\rho^*$ is continuously decreasing on $[0,1]$.
    \item $f$ is continuously increasing on $[\tau_1^*,\tau_0^*]$. 
    \item $\forall s_0\in [\tau_1^*,\tau_0^*]$ the sequence $s_{k+1}=f(s_k)$ converges to $\tau^*$.
\end{enumerate}

\textit{Proof of} (i). 
Define the synthetic gradients $\overline g$ and $g_\rho$ as
$$
\overline g(\tau)\triangleq J'(\tau)-\frac{1+\rho(\tau)}{2\tau}J(\tau),\qquad g_\rho(\tau)\triangleq J'(\tau)-\frac{1+\rho}{2\tau}J(\tau).
$$
We remark that
$\overline{\mathcal{C}}'(\tau)=0\Leftrightarrow\overline g(\tau)=0 $, and that $\mathcal{C}_\rho'(\tau)=0\Leftrightarrow g_\rho(\tau)=0$, where the first equivalence follows from the identity $J'(\tau)(1-\rho(\tau))=-J(\tau)\rho'(\tau)$.
As a consequence, $\tau^*$ and $\tau_\rho^*$ are the unique zeros of $\overline g$ and $g_\rho$ respectively, by \Cref{ass:unique_maximizers}.
Finally we obtain that $\tau=f(\tau)\Leftrightarrow\tau=\tau^*$, because $g_{\rho(\tau)}$ coincides with $\overline g$ and $\tau^*$ is its unique zero. \newline

\textit{Proof of} (ii).
Let
$\rho\in[0,1]$.  From \Cref{ass:unique_maximizers}, we deduce that the map $\psi(s)\triangleq2 s J'(s)/J(s)-1$  is continuously decreasing in a neighborhood of $\tau_\rho^*$. This observation follows from the expression of $g_\rho$ defined above. As a consequence, $\psi$ is locally invertible and the map $u\mapsto\psi^{-1}(u)=\tau_u^*$ is continuously decreasing in a neighborhood of $\rho$. \newline

\textit{Proof of} (iii). The claim follows from noticing that
$f:s\mapsto \psi^{-1}( \rho (s))$ is a composition of two continuously decreasing functions, from (ii) and 
\Cref{ass:decreasing_rho}.\newline

\textit{Proof of} (iv). We first show that $\tau_1^*< f(\tau_1^*)< \tau^*< f(\tau_0^*)< \tau_0^*$.   Indeed, since $\tau^*=\tau_{\rho(\tau^*)}$ and $\rho(\tau^*)\in(0,1)$, we obtain from (ii) that $\tau_1^*< \tau^*< \tau_0^*$. Using \Cref{ass:decreasing_rho}, we deduce that $1> \rho(\tau^*_1)> \rho(\tau^*)> \rho(\tau^*_0)> 0$. We conclude by applying (ii) to these last inequalities. 

From (i) and (iii), we know that $\tau^*$ is the unique fixed point of the continuous map $f$. Therefore, a similar inequality holds for any intermediate points: for any $s,s'\in[\tau_1^*,\tau_0^*]$ such that $s< \tau^*<s'$, we have $s< f(s)< \tau^*< f(s')< s'$.

Now, let $s_0\in[\tau_1^*,\tau^*)$ and $s_{k+1}=f(s_k)$. We obtain from (i) and (iii) that $s_k<f(s_k)=s_{k+1}< f(\tau^*)=\tau^*$. The sequence $(s_k)$ is increasing and upper bounded, thus it is convergent. The claim follows from (i) since the limit is necessarily a fixed point of $f$. The case $s_0\in(\tau^*,\tau_1^*]$ is symmetric.

\section{VARIANCE REDUCTION}

In this section, we derive a quantification of the variance reduction proposed in \Cref{sec:integration_time}, asymptotically as $\tau\rightarrow+\infty$. We also illustrate numerically the variance reduction on the Hierarchical Linear Model; see \Cref{sec:experiments}.

\subsection{Asymptotic Quantification}
Similarly to the previous section, we note $(x_\tau,v_\tau)_{\tau\ge 0}$ the stationary solution of the Langevin SDE defined in \Cref{sec:partial_momentum_refreshment}. We also note that our analysis holds for any continuously differentiable function $\varphi:\R^d\mapsto\R$ such that each  expectation of this section exists.

From \Cref{sec:integration_time}, we recall the unbiased estimators of the derivative of ${\rm ESJD}_\varphi$, defined as
\begin{align*}
    \delta(x_{\tau},x_0,v_{\tau})
&=
\big(\varphi(x_{\tau})-\varphi(x_0)\big)\nabla\varphi(x_{\tau})^{\top}M^{-1}v_{\tau},\\
g(x_\tau,v_\tau,x_0,v_0)&=\frac{1}{2}\Big(\delta(x_\tau,x_0,v_\tau)+\delta(x_0,x_\tau,-v_0)\Big).
\end{align*}
\begin{assumption}\label{ass:l2_mixing}
For every $\psi,\omega\in\mathbb{L}_2(\Pi^*)$, we have $\E[\psi(x_\tau,v_\tau)\, \omega(x_0,v_0)]\rightarrow \E[\psi(x_0,v_0)]\cdot\E[\omega(x_0,v_0)]$ as $\tau\rightarrow+\infty$.
\end{assumption}
This assumption holds whenever the kinetic Langevin diffusion mixes in $\mathbb{L}_2$; see \Cref{sec:partial_momentum_refreshment} for references. In the next proposition, we show that choosing the estimator $g$ instead of $\delta$ leads to a variance reduction of a factor $1/2$ as $\tau\rightarrow+\infty$. Studying the speed of convergence towards this factor is beyond our work, although we expect that an exponential decay can be derived under stronger assumptions, provided that these ensure geometric $\mathbb{L}_2$-mixing of the Langevin diffusion.\newline

\begin{proposition}\label{prop:variance_reduction}
 Suppose \Cref{ass:l2_mixing} holds. Then $\delta(x_\tau,x_0,v_\tau)$ and $\delta(x_0,x_\tau,-v_0)$ are asymptotically uncorrelated as $\tau\rightarrow+\infty$. Therefore
$$
\lim_{\tau\rightarrow+\infty}\frac{\V\big(g(x_\tau,v_\tau,x_0,v_0)\big)}{\mathbb{V}\big(\delta(x_\tau,x_0,v_\tau)\big)}\,=\,\frac{1}{2}.
$$
\end{proposition}

Since the estimators $\delta(x_\tau,x_0,v_\tau)$ and $\delta(x_0,x_\tau,-v_0)$ have the same distribution, we remark that
\[
\V\big(g(x_\tau,v_\tau,x_0,v_0)\big)=\frac{1}{2}\,\mathbb{V}\big(\delta(x_\tau,x_0,v_\tau)\big)+\frac{1}{2}\,\mathrm{Cov}\big(\delta(x_\tau,x_0,v_\tau),\delta(x_0,x_\tau,-v_0)\big).
\]
The claim of \Cref{prop:variance_reduction} follows by showing that the second term vanishes. We prove that, as $\tau\rightarrow+\infty$, we have:
\begin{enumerate}[label=(\roman*)]
    \item $\mathbb{E}\big[\delta(x_\tau,x_0,v_\tau)\big]\,\rightarrow\, 0$.
    \item $\mathbb{E}\big[\delta(x_\tau,x_0,v_\tau)\,\delta(x_0,x_\tau,-v_0)\big]\,\rightarrow\,0$.
\end{enumerate}

\textit{Proof of} (i). We make repeated use of the fact that $v_0$ is centered and independent from $x_0$. As a consequence, for every $\Pi$-integrable function $f:\R^d\rightarrow\R^d$, we have $\mathbb{E}[f(x_0)^\top M^{-1}v_0]=\mathbb{E}[f(x_0)]^{\top}M^{-1}\mathbb{E}[v_0]=0$. We combine this property with  \Cref{ass:l2_mixing}, and obtain that
\begin{align*}
    \lim_{\tau\rightarrow+\infty}\mathbb{E}\left[\delta(x_\tau,x_0,v_\tau)\right]&\quad=\quad
\lim_{\tau\rightarrow+\infty}\Big(\mathbb{E}\left[\varphi(x_\tau)\nabla\varphi(x_\tau)^\top M^{-1}v_\tau\right]-\mathbb{E}\left[\varphi(x_0)\nabla\varphi(x_\tau)^\top M^{-1}v_\tau\right]\Big)
\\
&\quad=\quad
\mathbb{E}\Big[\big(\varphi(x_0)\nabla\varphi(x_0)\big)^\top M^{-1}v_0\Big]-\mathbb{E}\Big[\varphi(x_0)\Big]\cdot\E\Big[\nabla\varphi(x_0)^\top M^{-1}v_0\Big]\quad=\quad0.
\end{align*}

\textit{Proof of} (ii). The time-reversibility of the Langevin dynamics implies that $(x_\tau,v_\tau,x_0,v_0)$ and $(x_0,-v_0,x_\tau,-v_\tau)$ have the same distribution; see \Cref{sec:integration_time}. Therefore
\begin{align*}
\mathbb{E}\big[\delta(x_\tau,x_0,v_\tau)\, \delta(x_0,x_\tau,-v_0)\big]\quad=&\quad
\mathbb{E}\left[\left(\varphi(x_{\tau})-\varphi(x_0)\right)^2\left(\nabla\varphi(x_{\tau})^{\top}M^{-1}v_{\tau}\right)\left(\nabla\varphi(x_{0})^{\top}M^{-1}v_{0}\right)\right]
\\
\quad=&\quad 2\,\mathbb{E}\Big[\varphi(x_{\tau})^2\left(\nabla\varphi(x_{\tau})^{\top}M^{-1}v_{\tau}\right)\left(\nabla\varphi(x_{0})^{\top}M^{-1}v_{0}\right)\Big]\\
\quad+&\quad 2\,\mathbb{E}\Big[\varphi(x_\tau)\varphi(x_{0})\left(\nabla\varphi(x_{\tau})^{\top}M^{-1}v_{\tau}\right)\left(\nabla\varphi(x_{0})^{\top}M^{-1}v_{0}\right)\Big].
\end{align*}
Again, we combine \Cref{ass:l2_mixing} with the fact that $v_0$ is centered and independent from $x_0$. We obtain that the two last terms vanish as $\tau\rightarrow+\infty$, since
\begin{align*}
  &\mathbb{E}\Big[\varphi(x_{\tau})^2\left(\nabla\varphi(x_{\tau})^{\top}M^{-1}v_{\tau}\right)\left(\nabla\varphi(x_{0})^{\top}M^{-1}v_{0}\right)\Big]\,\rightarrow\,
\mathbb{E}\Big[\varphi(x_{0})^2\left(\nabla\varphi(x_{0})^{\top}M^{-1}v_{0}\right)\Big]\cdot \mathbb{E}\Big[\nabla\varphi(x_{0})^{\top}M^{-1}v_{0}\Big]\,=\,0,\\
&\mathbb{E}\Big[\varphi(x_\tau)\varphi(x_{0})\left(\nabla\varphi(x_{\tau})^{\top}M^{-1}v_{\tau}\right)\left(\nabla\varphi(x_{0})^{\top}M^{-1}v_{0}\right)\Big]\,\rightarrow\,
\Big(\mathbb{E}\big[\varphi(x_0)\left(\nabla\varphi(x_{0})^{\top}M^{-1}v_{0}\right)\big]\Big)^2\,=\,0.
\end{align*}

\subsection{Numerical Illustration}
\begin{figure*}[!ht]
  \centering
    \includegraphics[scale=0.70]{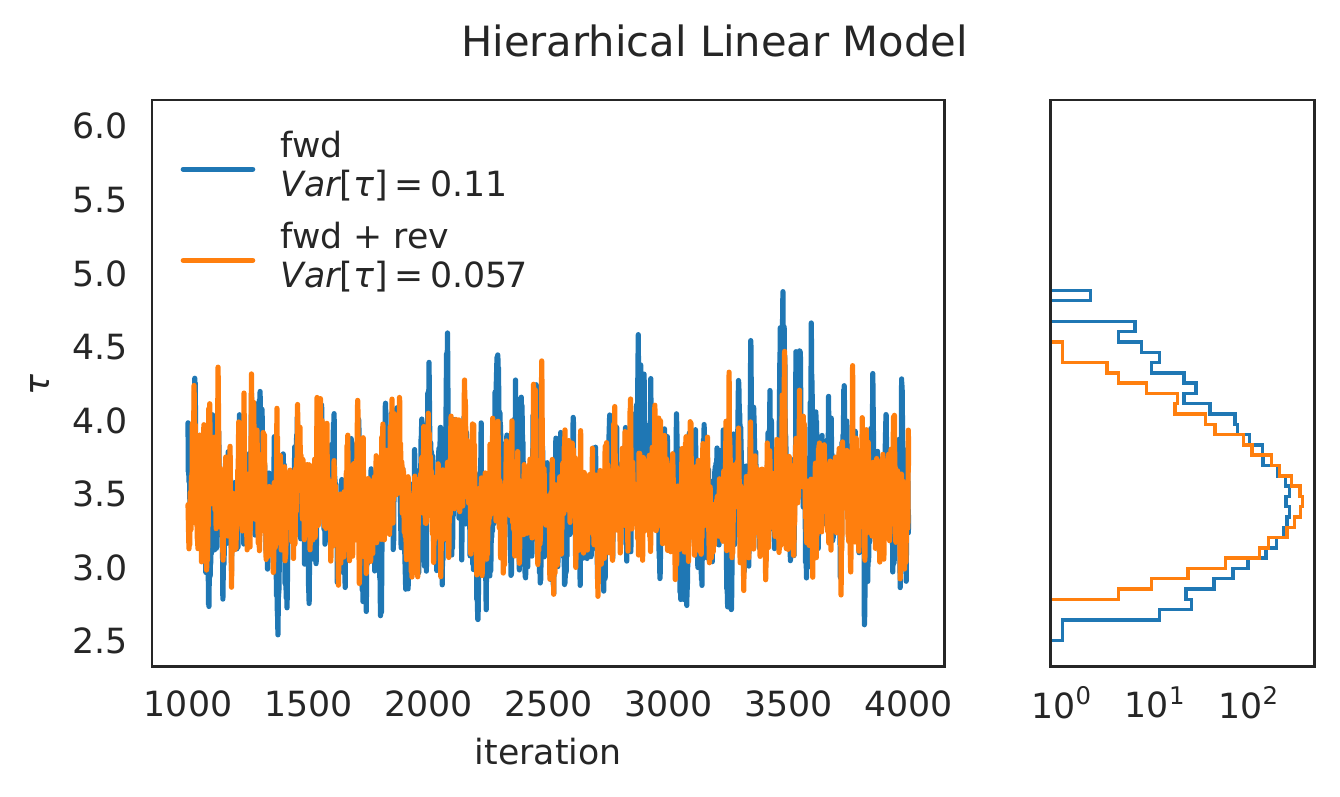}
\caption{
    Trace of the $\tau$ hyperparameter near convergence from two runs of Adaptive MALT on the Hierarchical Linear Model; see \Cref{sec:experiments}.
}
    \label{fig:variance}
\end{figure*}

An example trace of $\tau$ over time using the estimators $\delta$ (fwd), and  $g$ (fwd+rev), can be seen in \Cref{fig:variance}. The variation in the trace is caused primarily by the noise in the gradient estimate, and is reduced approximately by 2 when the improved estimator is used.

\clearpage
\section{RAW ESS VALUES}

\begin{figure*}[!tb]
  \centering
    \includegraphics[scale=0.65]{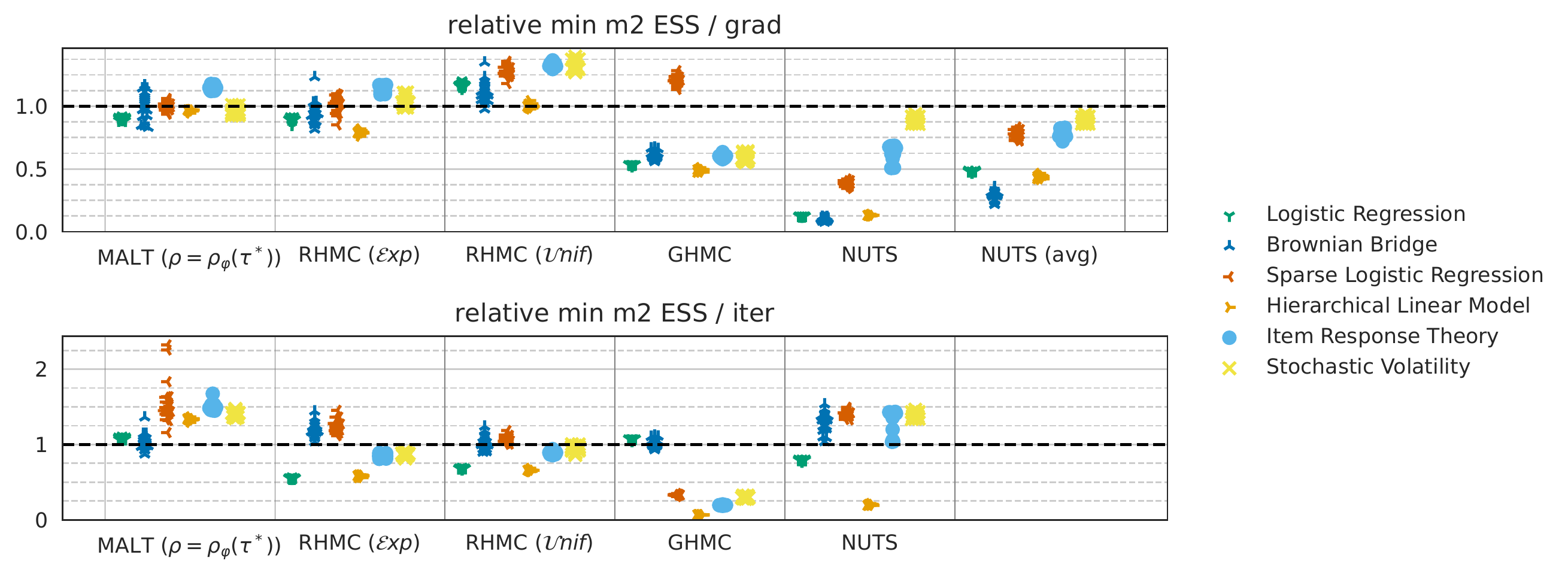}
\caption{
    Normalized run values used to derive \Cref{fig:ess}. Each configuration shows one point per run. Horizontal jitter added for clarity.
}
\label{fig:raw_ess}
\end{figure*}

In this section, we show the raw ESS/grad (\Cref{tab:ess_grad}) and ESS/iter (\Cref{tab:ess_iter}) numbers used to build \Cref{fig:ess}; see \Cref{sec:experiments}. We also show the raw run values in \Cref{fig:raw_ess}. We use the shorthand notation $\rho^*\triangleq\rho_\varphi(\tau^*)$.

\begin{table*}[!htb]

\caption{
    Minimum ESS/grad for estimating the centered second moment across algorithms and models. Values are 10th percentile across 20 runs.
    \vspace{0.2cm}
}
\label{tab:ess_grad}

\centering

\begin{tabular}{ llllllll }
\toprule
 & \makecell[l]{MALT \\ ($\rho=\rho^*$)} & \makecell[l]{MALT \\ ($\rho=1$)} & \makecell[l]{RHMC \\ ($\mathcal{E}xp$)} & \makecell[l]{RHMC \\ ($\mathcal{U}nif$)} & GHMC & NUTS & \makecell[l]{NUTS \\ (avg)} \\
\midrule
Logistic Regression & 9.98e-2 & 1.10e-1 & 9.81e-2 & 1.30e-1 & 5.88e-2 & 1.36e-2 & 5.31e-2 \\
Brownian Bridge & 1.06e-2 & 1.15e-2 & 1.08e-2 & 1.30e-2 & 7.14e-3 & 1.02e-3 & 3.06e-3 \\
Sparse Logistic Regression & 5.34e-3 & 5.32e-3 & 5.20e-3 & 6.85e-3 & 6.45e-3 & 1.98e-3 & 4.04e-3 \\
Hierarchical Linear Model & 5.61e-3 & 5.76e-3 & 4.57e-3 & 5.81e-3 & 2.80e-3 & 7.50e-4 & 2.48e-3 \\
Item Response Theory & 8.97e-3 & 7.72e-3 & 8.81e-3 & 1.04e-2 & 4.66e-3 & 4.10e-3 & 5.86e-3 \\
Stochastic Volatility & 1.23e-2 & 1.28e-2 & 1.31e-2 & 1.69e-2 & 7.42e-3 & 1.13e-2 & 1.13e-2 \\
\bottomrule
\end{tabular}
\end{table*}

\begin{table*}[!htb]

\caption{
    Minimum ESS/iter for estimating the centered second moment across algorithms and models. Values are 10th percentile across 20 runs.
    \vspace{0.2cm}
}
\label{tab:ess_iter}

\centering

\begin{tabular}{ lllllll }
\toprule
 & \makecell[l]{MALT \\ ($\rho=\rho^*$)} & \makecell[l]{MALT \\ ($\rho=1$)} &  \makecell[l]{RHMC \\ ($\mathcal{E}xp$)} & \makecell[l]{RHMC \\ ($\mathcal{U}nif$)}  & GHMC & NUTS \\
\midrule
Logistic Regression & 5.99e-1 & 4.78e-1 & 3.05e-1 & 3.77e-1 & 5.88e-1 & 4.36e-1 \\
Brownian Bridge & 7.20e-2 & 6.88e-2 & 8.33e-2 & 7.02e-2 & 7.14e-2 & 8.24e-2 \\
Sparse Logistic Regression & 2.69e-1 & 1.91e-1 & 2.33e-1 & 2.06e-1 & 6.45e-2 & 2.77e-1 \\
Hierarchical Linear Model & 5.61e-1 & 4.20e-1 & 2.43e-1 & 2.78e-1 & 2.80e-2 & 8.31e-2 \\
Item Response Theory & 3.59e-1 & 2.39e-1 & 2.02e-1 & 2.13e-1 & 4.66e-2 & 2.60e-1 \\
Stochastic Volatility & 3.56e-1 & 2.55e-1 & 2.21e-1 & 2.33e-1 & 7.42e-2 & 3.51e-1 \\
\bottomrule
\end{tabular}

\end{table*}

\section{HYPERPARAMETERS}

In this section we list the main hyperparameters used in experiments. For detailed implementation details, please refer to the source code available at \url{https://github.com/tensorflow/probability/tree/main/discussion/adaptive_malt}.

\begin{table*}[!htb]

\caption{
    Hyperparameters used in experiments.
    \vspace{0.2cm}
}
\label{tab:hyperparameters}

\centering

\begin{tabular}{ lll }
\toprule
 Hyperparameter & Meaning & Value \\
\midrule
$K$ & Number of parallel chains & 128 \\
$N_{clip}$ & Number of MALA steps & 100 \\
$N_{adapt}$ & Number of adaptation steps & 5000 \\
$\alpha^*$ & Target acceptance rate & 0.8 (0.7 NUTS) \\
$\eta$ & ADAM learning rate for $\log(h)$ and $\log(\tau)$ & 0.05 \\
$a$ & Learning rate for online estimates except for $w$ & $8$ \\
$a_w$ & Learning rate for $w$ & $3$ \\
$\beta_1$ & ADAM hyperparameter for $\log(\tau)$ & 0 \\
$\beta_2$ & ADAM hyperparameter for $\log(\tau)$ & 0.95 \\
$\epsilon$ & ADAM hyperparameter for $\log(\tau)$ & $10^{-8}$ \\
$N_{ADVI}$ & Number of ADVI learning steps & 2000 \\
$\eta_{ADVI}$ & ADVI learning learning rate & 0.1 \\
\bottomrule
\end{tabular}

\end{table*}

\end{appendices}
\end{document}